\documentclass{elsart}

\usepackage{graphicx}
\usepackage{multirow}
\usepackage{rotating}
\usepackage{wrapfig}
\graphicspath{{ps}}
\usepackage[usenames]{color}

\begin{document}

\begin{frontmatter}

\title{Study of $\bf \Omega_c^0$ and $\bf \Omega_c^{*0}$ Baryons at Belle}

\collab{Belle Collaboration}
   \author[ITEP]{E.~Solovieva}, 
   \author[ITEP]{R.~Chistov}, 
   \author[KEK]{I.~Adachi}, 
   \author[Tokyo]{H.~Aihara}, 
   \author[BINP]{K.~Arinstein}, 
   \author[Lausanne,ITEP]{T.~Aushev}, 
   \author[Sydney]{A.~M.~Bakich}, 
   \author[ITEP]{V.~Balagura}, 
   \author[JSI]{U.~Bitenc}, 
   \author[BINP]{A.~Bondar}, 
% ...

\address[BINP]{Budker Institute of Nuclear Physics, Novosibirsk, Russia}
% ...
\address[KEK]{High Energy Accelerator Research Organization (KEK), Tsukuba, Japan}
% ...
\address[ITEP]{Institute for Theoretical and Experimental Physics, Moscow, Russia}
\address[JSI]{J. Stefan Institute, Ljubljana, Slovenia}
% ...
\address[Lausanne]{\'Ecole Polytechnique F\'ed\'erale de Lausanne, EPFL, Lausanne, Switzerland}
% ...
\address[Sydney]{University of Sydney, Sydney, NSW, Australia}
% ...
\address[Tokyo]{Department of Physics, University of Tokyo, Tokyo, Japan}
% ...

\begin{abstract}
We report results from a study of the charmed double strange baryons $\Omega_c^0$ and  $\Omega_c^{*0}$ at Belle. The $\Omega_c^0$ is reconstructed using the $\Omega_c^0\to\Omega^-\pi^+$ decay mode, and its mass is measured to be $\left( 2693.6 \pm 0.3 {+1.8 \atop -1.5} \right)$ MeV/$c^2$. The $\Omega_c^{*0}$ baryon is reconstructed in the $\Omega_c^0\gamma$ mode. The mass difference $M_{\Omega_c^{*0}} - M_{\Omega_c^0}$ is measured to be $\left( 70.7 \pm 0.9 {+0.1\atop-0.9} \right)$ MeV/$c^2$.  The analysis is performed using 673 fb$^{-1}$ of data on and near the $\Upsilon$(4$S$) collected with the Belle detector at the KEKB asymmetric-energy $e^+e^-$ collider.
\end{abstract}

\end{frontmatter}

\section{Introduction}
The experimental study of charmed baryons, including measurement of their masses, widths and decay modes, is an important test of many theoretical models that provide quantitative predictions for the properties of heavy hadrons.

Unlike the $\Lambda_c^+, \Sigma_c^{++,+,0}, \Xi_c^{+,0}$ and even their excited states, a thorough experimental study of the charmed double strange baryon $\Omega_c^0$ is long overdue. The $\Omega_c^0$ $(J^P=({\frac{1}{2}})^+)$ is the heaviest known singly charmed hadron that decays weakly. The quark content of the $\Omega_c^0$ is {\it c$\{$ss$\}$}, where the {\it ss} pair is in a symmetric state. There are many theoretical models that predict the mass of the $\Omega_c^0$. However, the range of these predictions is rather wide: 2610--2786 MeV/$c^2$ \cite{Samuel}. Among several published measurements of the $\Omega_c^0$ mass only three report a statistically significant signal: E687 obtains $\left( 2699.9 \pm 1.5 \pm 2.5 \right)$ MeV/$c^2$ in the $\Omega_c^0\to\Sigma^+K^-K^-\pi^+$ channel~\cite{Frabetti1}, using five decay modes CLEO reports $\left( 2694.6 \pm 2.6 \pm 1.9 \right)$ MeV/$c^2$~\cite{Cronin}, BaBar quotes $\left( 2693.3 \pm 0.6 \right)$ MeV/$c^2$ in their analysis of $\Omega_c^{*0} \to \Omega_c^0 \gamma$~\cite{Aubert}. The world-average value of the $\Omega_c^0$ mass is $\left( 2697.5 \pm 2.6 \right)$ MeV/$c^2$~\cite{Yao}.

The decay mode $\Omega_c^0 \to \Omega^- \pi^+$ was observed for the first time by E687~\cite{Frabetti}. They reported $10.3 \pm 3.9$ events over the background of 5.8 events. With 13.7 fb$^{-1}$ of data CLEO also found a signal of $13.3 \pm 4.1$ events in this channel~\cite{Cronin}. FOCUS observes $23 \pm 7$ events of this decay mode~\cite{Link}. With a data sample of 230.7 fb$^{-1}$ BaBar reports a raw signal of $156 \pm 15$ events~\cite{Aubert}. 

The spin excited $J^P=({\frac{3}{2}})^+$ $\Omega_c^{*0}$ state was observed by BaBar only recently~\cite{Aubert}. They measured the difference between the masses of the $\Omega_c^{*0}$ and $\Omega_c^0$ to be $(70.8 \pm 1.0 \pm 1.1)$ MeV/$c^2$. Different QCD-based models yield a rather wide range of predictions for the mass difference, from 50 MeV/$c^2$ to 70 MeV/$c^2$~\cite{Mathur}. A lattice QCD calculation finds the value $\left( 94 \pm 10 \right)$ MeV/$c^2$~\cite{Voloshyn}.
 
In this work we present our results on the measurement of the $\Omega_c^0$ mass using the decay $\Omega_c^0\to\Omega^-\pi^+$ \cite{CC}. We also confirm the recent BaBar observation of $\Omega_c^{*0}\to\Omega_c^0\gamma$ \cite{Aubert}.

\section{Data and selection criteria}

The data used for this analysis were taken on the $\Upsilon (4S)$ resonance and in the nearby continuum using the Belle detector at the $e^+e^-$ asymmetric-energy collider KEKB~\cite{KEKB}. The integrated luminosity of the sample is equal to 605 fb$^{-1}$ of on-resonance data and 68 fb$^{-1}$ of data below the $\Upsilon$(4$S$).

Belle is a general-purpose detector based on a 1.5 T superconducting solenoid; a detailed description can be found elsewhere \cite{Abashian}. Tracking is performed by a silicon vertex detector (SVD) composed of concentric layers of double-sided silicon strip detectors, and a 50 layer drift chamber (CDC). Particle identification for charged hadrons, important for the measurement of final states with kaons and/or protons, is based on the combination of energy loss measurements ({\it dE/dx}) in the drift chamber, time-of-flight measurements (TOF) and aerogel Cherenkov counter (ACC) information. For each charged particle, measurements from these three subdetectors are combined to form likelihood ratios in the range from 0 to 1,
$$
P(A/B)={\cal L}(A)/({\cal L}(A)+{\cal L}(B)),
$$
where ${\cal L}(A)$ and ${\cal L}(B)$ are the likelihood values assigned to the $K, \pi$ or $p$ identification hypothesis for a given track. From now on we use these likelihood ratios to identify protons, charged kaons and pions with criteria that have efficiencies greater than $85\%$ for $p/K/\pi$. The probability for each particle species to be misidentified as one of the other two is less than $12\%$. 

An electromagnetic calorimeter (ECL) is used to detect photons, which are reconstructed from isolated ECL clusters that have no corresponding charged track, and a shower shape that is consistent with that of a photon.

$\Lambda$ hyperons are reconstructed in the $\Lambda\to p\pi^-$ decay mode, fitting the {\it p} and $\pi$ tracks to a common vertex and requiring an invariant mass in a $\pm 4$ MeV/$c^2$ ($\approx 4\sigma$) interval around the nominal $\Lambda$ mass value~\cite{Yao}. We then apply the following cuts on the $\Lambda$ decay vertex:

\begin{itemize}
\item the difference $\Delta z_{\Lambda}$ in the  z-coordinate between the proton and pion at the decay vertex should satisfy $\Delta z_{\Lambda}<$2 cm, where the z-axis is parallel to the $e^+$ beam direction;
\item the distance between the vertex position and interaction point (IP) in the plane transverse to the beam direction ($\Delta r_{\Lambda}^{xy}$) should be greater than 0.5 cm;
\item the angle $\beta_{\Lambda}$, between the $\Lambda$ momentum vector and the vector pointing from the IP to the $\Lambda$ decay vertex, should satisfy cos $\beta_{\Lambda}>$ 0.0;
\item the vertex fit should have an acceptable $\chi^2$.
\end{itemize}
We apply only a weak cut on $\beta_{\Lambda}$ since the analysis relies on $\Lambda$'s emerging from a displaced vertex rather than from the IP.

\begin{figure}[t]
\centering
\includegraphics[width=7.2cm]{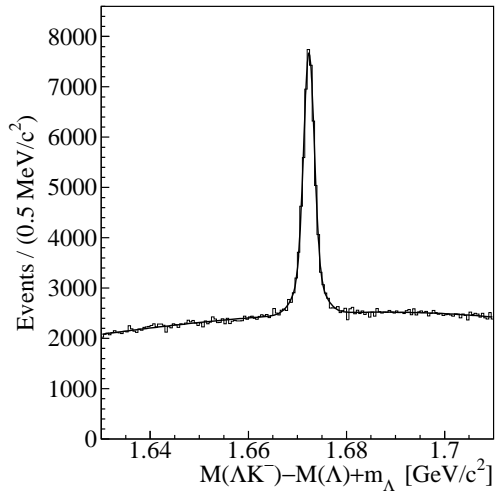}
\caption{$\Omega^-\to\Lambda K^-$: $M(\Lambda K^-) - M(\Lambda) + m_{\Lambda}$ spectrum of the selected $\Lambda K^-$ combinations. The selection requirements and the fit are described in the text.}
\label{f1}
\end{figure}

A sample of $\Omega^-$ hyperons is then reconstructed in the decay mode $\Omega^-\to\Lambda K^-$. The $K^-$ does not originate from  the  IP; we therefore impose only a loose requirement on the impact parameter of the kaon track, requiring it to be larger than $0.01$ mm. We then require the difference between the $\Lambda K^-$ invariant mass and the $\Lambda$ mass to be in the range 1669 MeV/$c^2 < M(\Lambda K^-) - M(\Lambda) + m_{\Lambda}< 1677$ MeV/$c^2$ ($\pm 3\sigma$), where $m_{\Lambda}$ is the nominal $\Lambda$ mass~\cite{Yao}, fit the $\Lambda$ and the $K^-$ track to a common vertex, and apply the following cuts:

\begin{itemize}
\item the distance in the transverse plane between the interaction point and the vertex position ($\Delta r_{\Omega^-}^{xy}$), obtained by fitting the $\Lambda$ trajectory and the $K^-$ track to a common point, should be greater than 0.01 cm;
\item the distance from the IP to the decay vertex for the $\Omega^-$ should be less than the corresponding distance for the $\Lambda$;
\item the $\Omega^-$ momentum vector should point to the IP, satisfying cos $\alpha_{\Omega^-}>$ 0.99, where the angle $\alpha_{\Omega^-}$ is measured between the $\Omega^-$ momentum vector and the vector pointing from the $\Omega^-$ decay vertex to the IP;
\item the angle $\alpha_{\Lambda}$, between the $\Lambda$ momentum vector and the vector pointing from the $\Omega^-$ to the $\Lambda$ decay vertex, should satisfy cos $\alpha_{\Lambda}>$ 0.99;
\item the $\Omega^-$ vertex fit should have an acceptable $\chi^2$.
\end{itemize}

Figure~\ref{f1} shows the $M(\Lambda K^-) - M(\Lambda) + m_{\Lambda}$ distribution for the $\Omega^-$ candidates after the requirements described above. A fit to this distribution with a double Gaussian to describe the signal and a third-order polynomial to describe the background gives 33880 $\pm$ 1680 $\Omega^-\to\Lambda K^-$ events. The fitted $\Omega^-$ mass of $\rm \left( 1672.363 \pm 0.012~(stat.) \right)$ MeV/$c^2$ is in good agreement with the PDG value \cite{Yao}.

\section{$\bf \Omega_c^0$ mass measurement}

Using the $\Omega^-$ sample we reconstruct $\Omega_c^0$ baryons in the decay mode $\Omega_c^0\to\Omega^-\pi^+$ without any requirement on their momenta to obtain $\Omega_c^0$ candidates from fragmentation of $c\bar c$ quarks as well as from $B$-meson decays.

\begin{figure}[bt]
\centering
\includegraphics[width=7.2cm]{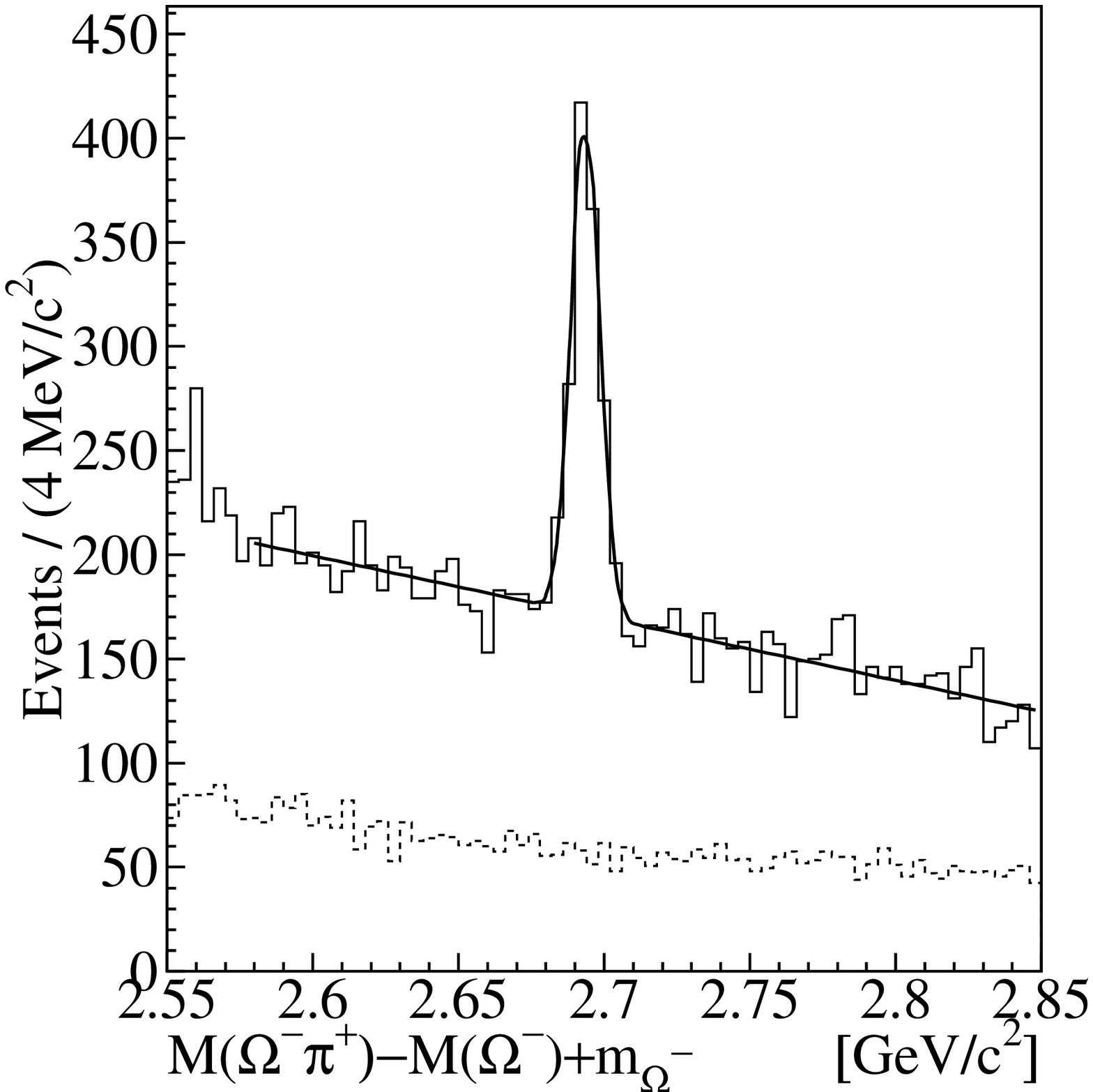}
\put(-55,150){\Large(a)}
\includegraphics[width=7.2cm]{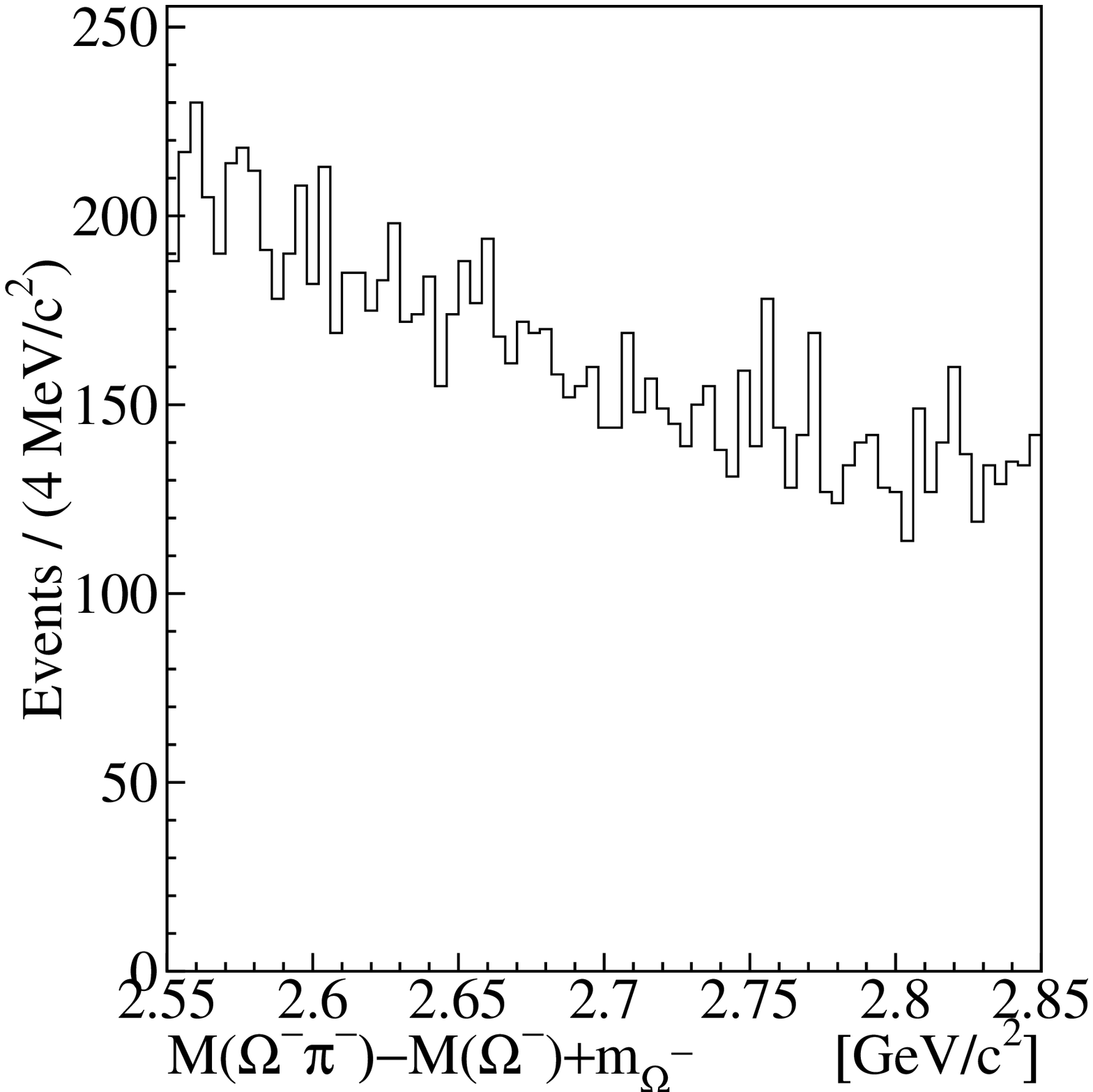}
\put(-55,150){\Large(b)}
\caption{(a): $\Omega_c^0\to\Omega^-\pi^+$: $M(\Omega^-\pi^+) - M(\Omega^-) + m_{\Omega^-}$ spectrum of the selected $\Omega^-\pi^+$ combinations. The selection requirements and the fit are described in the text. The dotted histogram shows the $\Omega^-$ sidebands. (b): $M(\Omega^-\pi^-) - M(\Omega^-) + m_{\Omega^-}$ spectrum for wrong sign $\Omega^-\pi^-$ combinations.}
\label{f2}
\end{figure}

Figure~\ref{f2}(a) shows the $M(\Omega^-\pi^+) - M(\Omega^-) + m_{\Omega^-}$ distribution for the $\Omega_c^0$ candidates: a clear signal peak is seen near 2700 MeV/$c^2$. The dotted histogram in Fig.~\ref{f2}(a) shows the contribution from the $\Omega^-$ sidebands, which is featureless. A fit ($\chi^2/n.d.f.=57.20/61$) to the distribution with a Gaussian for the signal and a first-order polynomial for the background yields 725 $\pm$ 45 events at a mass of $\left( 2693.6 \pm 0.3 \right)$ MeV/$c^2$; the Gaussian width was found to be $\left( 4.9 \pm 0.3 \right)$ MeV/$c^2$. We exclude the region to the left of 2580 MeV/$c^2$ from the fit to avoid a contribution from $\Omega_c^0\to\Omega^-\pi^+\pi^0$, where the $\pi^0$ is not reconstructed. To check the signal, we also reconstruct the wrong-sign combination $\Omega^-\pi^-$ (see Fig.~\ref{f2}(b)), where no peaking structures are observed, as expected. 

The systematic uncertainty on the measurement is assessed as follows. First we fit the signal in bins of $\Delta r_{\Omega^-}^{xy}$, $\Delta r_{\Lambda}^{xy}$, $\cos \alpha_{\Omega^-}$, $\cos \alpha_{\Lambda}$, $p^*(\Omega_c^0)$ and $p(\pi^+)$, where $p^*(\Omega_c^0)$ is the reconstructed momentum of the $\Omega_c^0$ candidate in the $e^+e^-$ center-of-mass frame and $p(\pi^+)$ is the $\pi^+$ reconstructed momentum in the laboratory frame. We divide the signal events into three roughly equally populated bins of the selected variables. In each pair of variables ($\Delta r_{\Omega^-}^{xy}$ and $\Delta r_{\Lambda}^{xy}$, $\cos \alpha_{\Omega^-}$ and $\cos \alpha_{\Lambda}$, $p^*(\Omega_c^0)$ and $p(\pi^+)$) we take the maximum value of the change of the fitted mass summed in quadrature, finding ${+1.8 \atop -1.5}$ MeV/$c^2$. We then vary the order of the polynomial in the fit function, fitting ranges and the Gaussian width within its error, finding no significant change in the fitted mass. Finally, since we actually measure $M(\Omega_c^0)-M(\Omega^-)+m_{\Omega^-}$, the contribution from the $\Omega^-$ mass uncertainty is $\pm 0.29$ MeV/$c^2$ \cite{Yao}. All the sources of systematic uncertainty are summarized in Table~\ref{t1}. We obtain:
$$
M_{\Omega_c^0} = \rm \left( 2693.6 \pm 0.3~(stat.)~{\textstyle {+1.8\atop -1.5}}~(syst.)\right)~MeV/{\it c}^2.
$$
Our measurement of the $\Omega_c^0$ mass is in agreement with the world-average value~\cite{Yao}.

\section{Confirmation of the $\bf \Omega_c^{*0}$}

\begin{figure}[t]
\centering
\includegraphics[width=7.2cm]{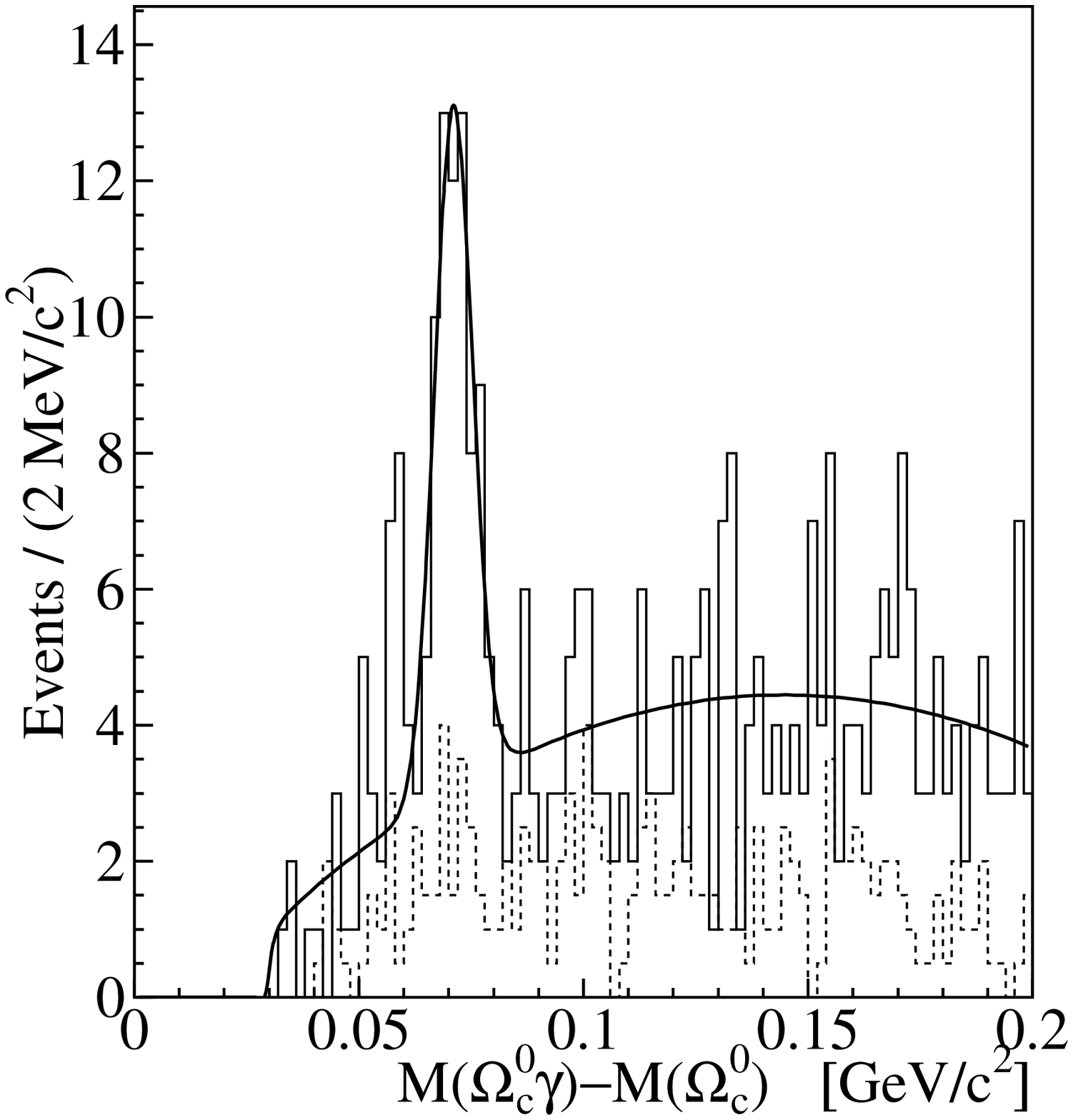}
\put(-55,150){\Large(a)}
\includegraphics[width=7.2cm]{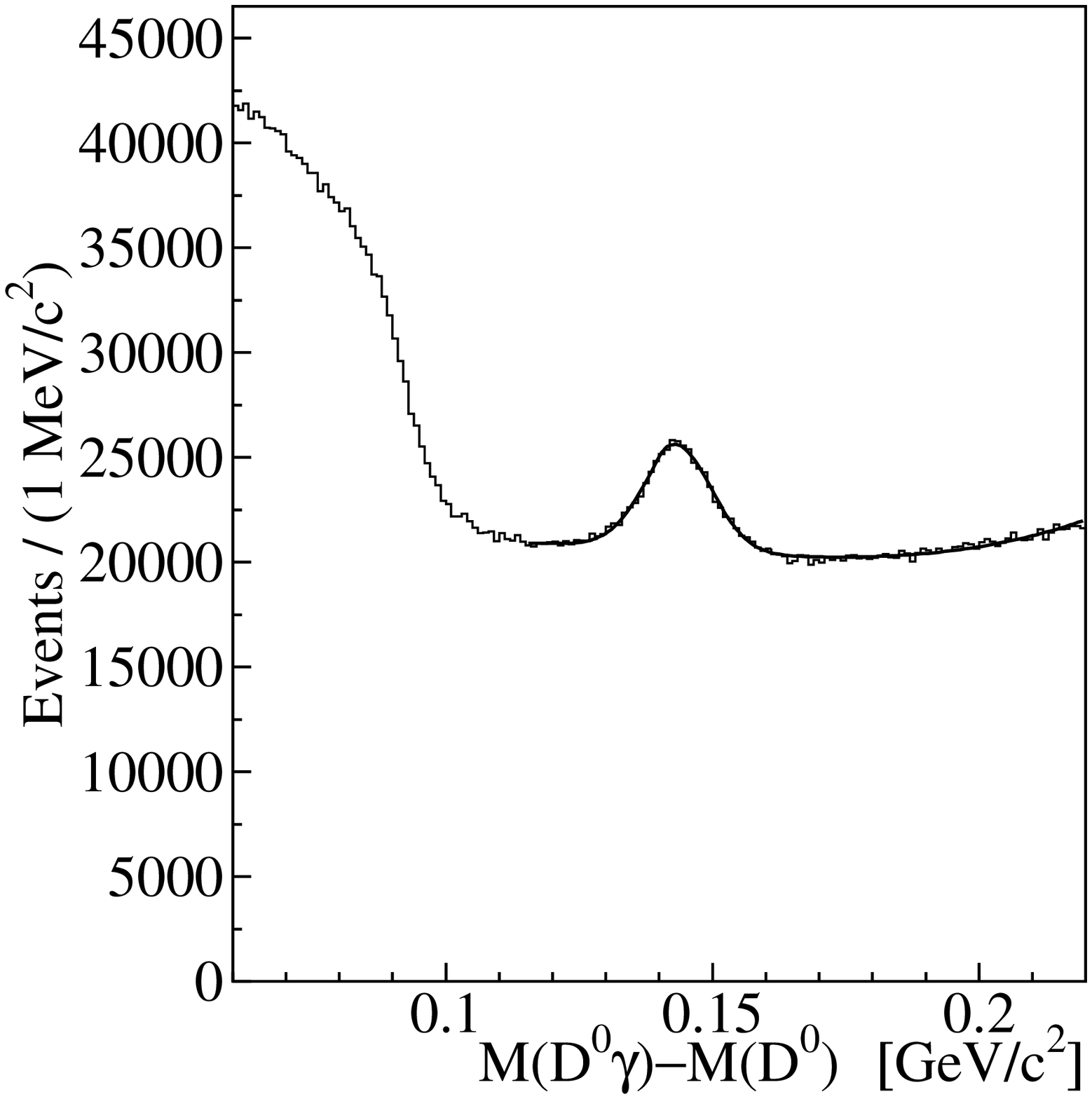}
\put(-55,150){\Large(b)}
\caption{ (a): $\Omega_c^{*0}\to\Omega_c^0\gamma$ ($\Omega_c^0\to\Omega^-\pi^+$): $M(\Omega_c^0\gamma) - M(\Omega_c^0)$ spectrum of the selected $\Omega_c^0\gamma$ combinations. The dotted histogram shows the $\Omega_c^0$ sidebands. (b): $D^{*0}\to D^0\gamma$ ($D^0\to K^-\pi^+$): $M(D^0\gamma)-M(D^0)$ spectrum of the selected $D^0\gamma$ combinations. The selection requirements for (a) and (b) as well as the fit are described in the text.}
\label{f4}
\end{figure}

From the $\Omega_c^0$ sample we reconstruct the first excited $\Omega_c^0$ baryon in the decay mode $\Omega_c^{*0}\to\Omega_c^0\gamma$. We apply a 2679 MeV/$c^2 < M(\Omega_c^0) <$ 2709 MeV/$c^2$ requirement on the $\Omega_c^0$ candidate. To suppress the combinatorial background, we require the photon candidate to have energy $E_{\gamma} > 100$ MeV and $\Omega_c^{*0}$ candidates to have momenta $p^* > 2.5$ GeV/$c$, where $E_{\gamma}$ is the energy detected in the electromagnetic calorimeter and $p^*$ is the reconstructed momentum of the $\Omega_c^{*0}$ candidate in the $e^+e^-$ center of mass.

Figure~\ref{f4}(a) shows the mass difference $M(\Omega_c^0\gamma) - M(\Omega_c^0)$ distribution for the remaining $\Omega_c^0\gamma$ candidates: a clear signal peak in the mass difference $M(\Omega_c^0\gamma) - M(\Omega_c^0)$ is seen near 70 MeV/$c^2$. The dotted histogram in Fig.~\ref{f4}(a) represents the $\Omega_c^0$ sidebands, which are featureless. A fit ($\chi^2/n.d.f.= 70.69/73$) to the distribution with a Crystal Ball function~\cite{Skwarnicki} with parameters fixed from the Monte Carlo simulation for the signal and a second-order polynomial with an arctangent threshold term for the background yields 54 $\pm$ 9 events at a mass difference of $\left( 70.7 \pm 0.9 \right)$ MeV/$c^2$. The fit gives a significance of $6.4\sigma$ for the signal including systematics from varying the signal and background parameterizations.

The systematic uncertainty on the $M(\Omega_c^0\gamma)-M(\Omega_c^0)$ measurement is assessed as follows. We vary the order of the polynomial and the threshold term in the fit function and signal width within its error, finding a mass shift of $+0.1$ MeV/$c^2$ and $+0.1 \atop -0.2$ MeV/$c^2$ respectively (Table~\ref{t1}).

We study a possible bias in the measurement from the incomplete calibration of detector responses (energy scale and line shape) for low energy photons. Comparing Monte Carlo generated and reconstructed masses for $\Omega_c^{*0}$ we find a shift in the mass of $+0.9$ MeV/$c^2$. As a cross check, we reconstruct $D^{*0}\to D^0\gamma$, $D^0\to K^-\pi^+$ using the full data set. Besides the particle identification selection described in Section 2, we require the $D^0$-meson invariant mass to be within $\pm 10$ MeV/$c^2$ of the nominal value ($\sim \pm 2 \sigma$), the photon to have energy $E_{\gamma} > 100$ MeV, and the $D^{*0}$ candidate to have momentum $p^* > 2.5$ GeV/$c$. By analysing the $E_{\gamma}$ spectrum from the 50~MeV/$c^2<M(\Omega_c^0\gamma) - M(\Omega_c^0)<90$ MeV/$c^2$ mass window we find 82\% of all $\gamma$'s are in the energy range from 100 to 200 MeV. Therefore we require the photons from the $D^{*0}$ decays to lie in the same energy range. Figure~\ref{f4}b shows the $M(D^0\gamma)-M(D^0)$ mass distribution for the remaining $D^0\gamma$ candidates after these requirements. From a fit by a Crystal Ball function for the signal and a third-order polynomial for the background we obtain $M(D^0\gamma) - M(D^0) = \left( 143.03 \pm 0.09 \right)$ MeV/$c^2$. This is 0.91 MeV/$c^2$ higher than the world-average mass of $m_{D^{* 0}}-m_{D^0}$ \cite{Yao}, which is consistent with the Monte Carlo study. We assign a conservative systematic error of $-0.9$ MeV/$c^2$ due to the uncertainty of calibrations.

\begin{table}[t]
\centering
\caption{Contributions to the systematic uncertainty.}
\begin{tabular}{c|c|c|}
   \cline{2-3}
    & $M_{\Omega_c^0}$, & $M_{\Omega_c^{*0}} - M_{\Omega_c^0}$, \\
    & \multicolumn{1}{|c|}{MeV/$c^2$} & \multicolumn{1}{|c|}{MeV/$c^2$} \\
   \hline
   \multicolumn{1}{|c|}{Fit in bins} & $+1.8 \atop -1.5$ & - \\
   \multicolumn{1}{|c|}{$m_{\Omega^-}$~\cite{Yao}} & $\pm$0.3 & - \\
   \multicolumn{1}{|c|}{\footnotesize Calibration mode} & - & $-$0.9 \\
   \multicolumn{1}{|c|}{Signal width} & 0.0 & $+0.1 \atop -0.2$ \\
   \multicolumn{1}{|c|}{Fit function} & 0.0 & $+$0.1 \\
   \hline
   \multicolumn{1}{|c|}{Total} & $+1.8 \atop -1.5$ & ${+0.1\atop-0.9}$ \\
   \hline
   \hline
\end{tabular}
\label{t1}
\end{table}

Combining these results (see Table~\ref{t1}), we find
$$
M_{\Omega_c^{*0}} - M_{\Omega_c^0} = \rm \left( 70.7 \pm 0.9~(stat.)~{\textstyle {+0.1 \atop -0.9}}~(syst.) \right)~MeV/{\it c}^2.
$$
in good agreement with $\left( 70.8 \pm 1.0 \pm 1.1 \right)$ MeV/$c^2$ obtained in Ref.~\cite{Aubert}.
\section{Summary}

In conclusion, we report a precise measurement of the $\Omega_c^0$ mass using its decay to $\Omega^- \pi^+$, $M_{\Omega_c^0} = \rm \left(2693.6 \pm 0.3~(stat.)~{+1.8 \atop -1.5}~(syst.) \right)~MeV/{\it c}^2$. The presented analysis provides a measurement of the $\Omega_c^0$ mass based on the largest available statistics. The result agrees with the world-average value~\cite{Yao} and has a significantly smaller uncertainty.


We also search for the $\Omega_c^{*0}$ meson decaying into $\Omega_c\gamma$ that was recently observed by BaBar~\cite{Aubert}. We use our sample of $\Omega_c^0\to\Omega^- \pi^+$ decays and confirm the production of $\Omega_c^{*0}$ in $c\bar c$ fragmentation with a significance of $6.4\sigma$. The mass difference $M_{\Omega_c^{*0}} - M_{\Omega_c^0}$ is measured to be $\rm \left( 70.7 \pm 0.9~(stat.)~{+0.1 \atop -0.9}~(syst.) \right) MeV/{\it c}^2$.

\vspace{1.5 cm}

We thank the KEKB group for the excellent operation of the accelerator, the KEK cryogenics group for the efficient operation of the solenoid, and the KEK computer group and the National Institute of Informatics for valuable computing and SINET3 network support. We acknowledge support from the Ministry of Education, Culture, Sports, Science, and Technology of Japan and the Japan Society for the Promotion of Science; the Australian Research Council and the Australian Department of Education, Science and Training; the National Natural Science Foundation of China under contract No.~10575109 and 10775142; the Department of Science and Technology of India; the BK21 program of the Ministry of Education of Korea, the CHEP SRC program and Basic Research program (grant No.~R01-2005-000-10089-0) of the Korea Science and Engineering Foundation, and the Pure Basic Research Group program of the Korea Research Foundation; the Polish State Committee for Scientific Research; the Ministry of Education and Science of the Russian Federation and the Russian Federal Agency for Atomic Energy; the Slovenian Research Agency; the Swiss National Science Foundation; the National Science Council and the Ministry of Education of Taiwan; and the US Department of Energy.


\begin{thebibliography}{99}
\bibitem{CC} Charge-conjugate modes are implicitly included throughout this letter.
\end{thebibliography}
\end{document}